\documentclass[rme]{ipart_4up}

\Vol{1}
\Issue{1}
\Year{2026}
\Doi{10.4310/RME.260101666666}
\Pubonline{February 13, 2026}
\firstpage{1}

\usepackage{mathptmx}
\usepackage{amsthm}
\usepackage{yfonts}

\usepackage{bbm}
\usepackage{bm}
\usepackage{mathrsfs}
\usepackage{float}
\usepackage[most]{tcolorbox}  % 提供彩色盒子
\usepackage{xcolor}
\usepackage{lipsum}           % 示例文本，可删

\usepackage[colorinlistoftodos]{todonotes}

\usepackage{hyperref}
\hypersetup{
	colorlinks=true, % 启用链接颜色 
	linkcolor=blue, % 章节等内部链接的颜色 
	citecolor=blue, % 参考文献链接的颜色 
	urlcolor=blue % URL链接的颜色 
}

\newtcolorbox{promptbox}[1][]{
  enhanced, breakable,
  colback=gray!5,        % 内容背景
  colframe=black,        % 边框颜色
  boxrule=0.6pt,         % 边框粗细
  arc=2pt,               % 圆角
  left=1em, right=1em, top=0.7em, bottom=0.7em,
  colbacktitle=black,    % 标题条背景
  coltitle=white,        % 标题文字颜色
  fonttitle=\bfseries,   % 标题加粗
  title=#1,              % 可选参数作为标题
  attach boxed title to top left={yshift=-2mm}, % 标题贴顶
  boxed title style={frame code={}, interior code={\fill[black]
     (frame.south west) rectangle (frame.north east);}},
}

\newcommand{\be}[0]{\begin{equation}}
\newcommand{\ee}[0]{\end{equation}}

\numberwithin{equation}{section}

\theoremstyle{plain}% default

\begin{document}

\title[Textual semantics and machine learning methods for data product pricing]{Textual semantics and machine learning methods for data product pricing}

\author[R. Gao, F. Xiao, J. Li, and S. Cui]{Ruize Gao, Feng Xiao, Jinpu Li, and Shaoze Cui\corrauthor}
\corrauthortext{Corresponding author.}

%\footnote{Ruize Gao, Beijing Institute of Mathematical Sciences and Applications, Beijing, China. 
%Feng Xiao, Beijing Institute of Mathematical Sciences and Applications, Beijing, China. 
%Jinpu Li, Tsinghua University, Beijing, China. 
%Shaoze Cui, Corresponding author, Beijing Institute of Technology, Beijing, China, shaoze-cui@foxmail.com; shaoze-cui@bit.edu.cn}}

\begin{abstract}
Reasonable pricing of data products enables data trading platforms to maximize revenue and foster the growth of the data trading market. The textual semantics of data products are vital for pricing and contain significant value that remains largely underexplored.
Therefore, to investigate how textual features influence data product pricing, we employ five prevalent text representation techniques to encode the descriptive text of data products. And then, we employ six machine learning methods to predict data product prices, including linear regression, \linebreak{neural networks, decision trees, support vector machines, random forests, and XGBoost}. Our empirical design consists of two tasks: a regression task that predicts the continuous price of data products, and a classification task that discretizes price into ordered categories. Furthermore, we conduct feature importance analysis by the mRMR feature selection method and SHAP-based interpretability techniques. Based on empirical data from the AWA Data Exchange, we find that for predicting continuous prices, Word2Vec text representations capturing semantic similarity yield superior performance. In contrast, for price-tier classification tasks, simpler representations that do not rely on semantic similarity, such as Bag-of-Words and TF-IDF, perform better. SHAP analysis reveals that semantic features related to healthcare and demographics tend to increase prices, whereas those associated with weather and environmental topics are linked to lower prices. This analytical framework significantly enhances the interpretability of pricing models.

\medskip
\noindent\setkeywords

\end{abstract}

\maketitle

%\keywords{Data product price, Machine learning, Textual representation, Feature importance analysis}
%\setkeywords

% \subjclass{abc99d}
% \setsubjclass

\section{Introduction} \label{sec:introduction}

In the era of digital economy, data referred to as the “new oil” has become an invaluable resource \citep{pei2020survey,zhu2024dataprice}. Governments, enterprises, and individuals worldwide are increasingly leveraging data-driven approaches to unlock substantial economic value. For example, the U.S. healthcare sector could generate more than USD 300 billion in value annually by fully harnessing big data.\footnote{Big data: The next frontier for innovation, competition, and productivity, McKinsey Global Institute, https://www.mckinsey.com/capabilities/tech-and-ai/our-insights/big-data-the-next-frontier-for-innovation.} 
In the retail sector, the adoption of dynamic pricing algorithms based on consumer behavioral data has been shown to increase revenue by 11\% to 19\% \citep{fisherUsingDataBig2018}. Similarly, the UK’s HM Revenue \& Customs (HMRC) recovered GBP 1.4 billion in unpaid taxes in a single year through its Connect system, which integrates 28 data sources to support anomaly detection and risk modeling \citep{maciejewskiMoreBetterFaster2017}. The rapid growth in data-driven demand has fueled the rise of Data Marketplaces, which is an online platforms that bring together data providers, consumers, and brokers to facilitate the exchange of data products \citep{zhangSurveyDataPricing2023}. According to a report by Grand View Research (2025), the global DMs market was valued at approximately USD 1.49 billion in 2024 and is predicted to reach USD 5.73 billion by 2030.\footnote{Data Marketplace Platform Market (2025-2030), Grand View Research, https://www.grandviewresearch.com/industry-analysis/data-marketplace-market-report.}

In data marketplaces, an effective data pricing model can enhance transaction efficiency, enable platforms to maximize revenue, and foster the sustained growth of the market \citep{zhangSurveyDataPricing2023,pei2020survey}. Traditional cost-based pricing models often underestimate the intrinsic value embedded in data \citep{liConceptsAccountingTreatment2024,wangResearchPricingMethod2022}. Revenue-based pricing models, in contrast, may generate biased prices because different data consumers form heterogeneous expectations about future benefits, making it difficult to establish a fair and consistent price. In comparison, market-based and data-driven pricing approaches allow data providers to determine more reasonable prices by consider the features of the data itself and existing prices of comparable datasets. Such mechanisms further incentivize data providers to continuously improve data quality and proactively participate in data trade.

Existing data-driven approaches to data product pricing mainly include linear regression models, machine learning techniques, and explainable learning methods. In terms of linear regression approaches, regression models can capture the linear relationships between data product prices and their influencing factors. For example, \cite{han2022trading} employed Lasso regression to identify features that are valuable to users and then combined these features with seller-defined unit prices to enable automatic pricing. About machine learning (ML) methods, ML-based techniques help uncover nonlinear relationships between data features and data product prices. For instance, \cite{hao2024hceg} introduced a heterogeneous ensemble pricing model based on clustering strategies to enhance the accuracy of data asset valuation. \cite{shang2025lightgbm} applied a LightGBM-based model for pricing medical data. For explainable learning, existing studies primarily adopt methods such as SHAP to analyze the influence of different features on data product prices. For example, \cite{zhu2024dataprice} generated semantic embeddings of dataset descriptions using mBERT and then applied SHAP to evaluate the relative importance of pricing factors.

Although existing studies have made initial investigation in applying machine learning methods to data product pricing, most current work still relies primarily on numerical features (e.g., dataset size, update frequency, category) to characterize data products. Few studies extracts value-relevant information from textual semantics. In practice, on data marketplaces such as Amazon Data Exchange and Datarade, data products are typically presented through natural language descriptions, including data product titles, summaries, detailed descriptions, and application scenarios. The semantic content embedded in this text not only reflects the product’s thematic and domain attributes but also conveys implicit signals about its market positioning and perceived value \citep{ganKeywordsAreNot2025}. Although \cite{zhu2024dataprice} made an early attempt to incorporate textual features into data product pricing, their approach relied solely on a single mBERT-based textual representation. It did not consider other widely used textual representations such as Bag-of-Words (BOW), TF-IDF, or topic-based models. 

Current challenges in using textual data for data product pricing include the following aspects.
First, there exist multiple approaches to text representation, ranging from context-independent methods such as BOW and TF-IDF, to context-aware methods such as Word2Vec, as well as topic-model-based representations. However, it remains unclear which types of textual representations are most appropriate for characterizing data products and for supporting pricing decisions.
Second, data product pricing typically involves two main paradigms: tiered pricing and precise pricing. Tiered pricing enables rapid classification of data products into different price levels, thereby providing an approximate price range, while precise pricing focuses on determining an explicit price to serve as a basis for negotiation between data providers and data consumers. The respective roles and effectiveness of different text representation methods in these distinct pricing tasks have not yet been systematically investigated.

Therefore, to investigate how different textual feature representation methods influence various types of data product pricing tasks, we conduct a systematic empirical analysis. First, we extract textual information from data products and represent it using five prevalent textual feature engineering methods: Bag-of-Words, TF-IDF, Word2Vec, LDA, and BERTopic. These methods capture textual semantics from multiple perspectives and provide a comprehensive quantification of data product descriptions. Second, we integrate these textual features with structured attributes such as numerical and categorical variables, and construct pricing models using a diverse set of machine learning algorithms, including linear regression, neural networks, decision trees, support vector regression, random forests, and XGBoost. In the empirical analysis, we design two prediction tasks, a regression task for modeling the continuous prices of data products and a classification task that categorizes prices into discrete tiers, to thoroughly compare the performance of different textual representation methods across distinct prediction scenarios. Finally, we conduct feature-importance analysis using mRMR for feature selection and SHAP for model interpretability. These analyses identify the key features of data product pricing and further validate the critical role of textual semantic features in improving pricing prediction performance.

The rest of this study is organized as follows. The Section 2 reviews the related studies in the field of data pricing. The Section 3 presents the research methodology, including the representation of textual and numerical features, the prediction models, and the performance evaluation metrics. The Section 4 reports the empirical results, covering data description, comparative analyses of regression and classification predictions based on different feature representations, and feature importance analysis. The Section 5 discusses the theoretical and managerial implications. Finally, the Section 6 concludes the main findings.

\section{Literature Review} \label{sec:literature}

\subsection{A taxonomy of studies on data pricing}

Academic research on data pricing encompasses a range of perspectives. As such, it is important to clearly define both the subject of analysis and the perspective being used when addressing this topic. An influential study by \cite{pei2020survey} highlights that data pricing is an interdisciplinary issue, combining economic theory with knowledge from data science. Specifically, understanding the fundamental properties of data requires statistical modeling, while addressing pricing issues involves characterizing elements such as supply and demand relationships and market structure \citep{azcoitia2022survey}. Given this, we classify data pricing research into two main approaches: economic and data science perspectives.

Studies from an economic or management perspective primarily focuses on topics such as data market design, data sales models, and data asset valuation \citep{huang2021toward,bergemann2022economics,baley2025data}. Among these, game theory, auction, and mechanism design are widely used to study the behavior of data market participants and the rules of data trade \citep{bergemann2021information}. For example, \cite{markovich2025competing} discuss whether platforms choose to monetize user data as a business model under the influence of network effects. \cite{bhargava2025strategic} develop a model of cross-market competition, where data gathered from consumer activity in one market enhances product quality in another.

In contrast, data pricing from a data science perspective focuses more on valuation or pricing strategies in specific contexts. The first key issue is defining the object of pricing. For example, \cite{congDataPricingMachine2022} review the pricing objects within machine learning pipelines, including raw datasets, data labels, and machine learning models. The specific data structure and form of the pricing object determine the formulation of mathematical or statistical models. For example, when discussing pricing data queries, the no-arbitrage condition is a key constraint \citep{lin2014arbitrage}. In the context of collaborative machine learning, a key issue is how participants can receive a fair allocation based on their data contributions, and the improved Shapley value method provides a solution to this challenge \citep{jia2019towards}. For pricing datasets or data products in a general sense, factor models are typically used \citep{heckman2015pricing}. A simple and intuitive approach is to use linear models, similar to the hedonic model used in real estate valuation \citep{ekeland2004identification}. Linear models offer good interpretability, but they have limitations in terms of practical applicability for data pricing. As a result, more research has turned to machine learning approaches.

\subsection{Machine learning approaches to data pricing}
Initial machine learning applications in data pricing often extended traditional statistical methods, focusing on dynamic inputs and the predictive power of the model. Regression models, for example, are commonly used to fit utility functions or manage privacy-related costs. \cite{han2022trading} utilize LASSO regression to fit user demand against seller-defined unit prices, allowing for reasonable automatic pricing while effectively preventing overfitting. When dealing with personal data, pricing models must account for privacy sensitivity. \cite{xu2016dynamic} address this by applying Ridge regression within a Multi-Armed Bandit (MAB) framework, creating a dynamic pricing mechanism that balances revenue fluctuations. Similarly, \cite{xiao2023locally} use Stochastic Gradient Descent (SGD) to learn the data cost associated with privacy perturbations, thereby achieving a dual goal of profit maximization and user privacy protection.

To capture more complex, non-linear relationships in data pricing, many studies have adopted powerful ensemble learning models known for their high predictive accuracy. For instance, \cite{zhao2024research} uses 
an XGBoost model for prediction, pairing it with SHAP (SHapley Additive exPlanations) to maintain high accuracy while ensuring model interpretability. For pricing high-dimensional and non-linear data, such as in healthcare, \cite{shang2025lightgbm} leverage the capabilities of LightGBM. A distinct trend in this area involves creating novel ensemble strategies by combining multiple heterogeneous models. \cite{lv2025research} demonstrates that a stacking ensemble, which combines three base models and a fusion model, outperforms any individual model. \cite{hao2023ensemble} are among the first to apply a Rank-Prune-Average strategy, which averages predictions from the best-performing models in a heterogeneous set. Building on this, \cite{hao2024hceg} develop a more sophisticated fusion method that uses Random Forest (RF) for feature selection, K-means for clustering, and a ``gravity-based" weighting strategy to combine fifteen different models. 

Given that data products are often described by unstructured text, deep learning models from Natural Language Processing (NLP) have also become essential. These models are adept at embedding textual descriptions or detecting sensitive information. \cite{hu2025sddp} implement a BERT-Attention-GCN model, which uses BERT for text representation and a GCN to detect sensitive information, enabling user-controlled pricing. Another notable system. \cite{zhu2024dataprice} also use mBERT to embed dataset descriptions, which are then fed into RF and KNN models; this system uniquely allows users to input natural language descriptions and receive price range recommendations complete with interpretable, Shapley-based explanations.

\subsection{Contribution}
We contribute this data pricing field from the following three main aspects:

First, we incorporate multiple textual representation methods to predict data product prices, which extends existing data pricing research that primarily relies on numerical features and provides a meaningful enhancement from the perspective of textual semantics.

Second, we systematically compare the performance of different textual representation methods across distinct pricing tasks. Through a dual experimental design, including continuous price prediction and discrete price-tier classification, we examine how various text representations influence the modeling performance.

Finally, we analyze the importance of multidimensional features in data product price prediction by combining the mRMR method and SHAP values. And we propose a novel approach to map these embeddings back to informative textual keywords to reveal crucial influencing factors, which highlight the significant contribution of semantic embedding features.

\section{Methodology} \label{sec:methodology}

This section mainly consists of problem statement, feature representation, feature selection models, data price prediction models, and performance evaluation metrics.

\subsection{Problem statement}

In this study, the model inputs primarily consist of three types of features derived from data products: textual descriptions, numerical attributes, and categorical attributes.
For textual data, we employ multiple representation methods, including Bag-of-Words (BOW), Term Frequency–Inverse Document Frequency (TF-IDF), Word2Vec, Latent Dirichlet Allocation (LDA), and BERTopic.
For categorical data, we adopt one-hot encoding to transform categorical attributes into numerical feature vectors.
The modeling framework primarily relies on a range of machine learning methods, including Logistic Regression (LR), Artificial Neural Networks (ANN), Decision Trees (DT), Support Vector Machines (SVM), Random Forests (RF), and XGBoost.

The model outputs involve two tasks: a price-interval classification task and a point-price regression task. We include price-interval prediction as a complementary task to point-price regression for both economic and practical reasons. First, data are information goods whose realized value is buyer-specific: different buyers combine the same dataset with different existing data assets, models, and use cases, leading to substantial dispersion in willingness to pay \citep{farboodi2025valuing}. In such settings, a predicted range can be more actionable for screening and negotiation than a single point estimate. Second, data marketplaces commonly implement ``versioning” or tiered pricing and may adjust prices dynamically as more information about demand, usage, and product quality becomes available \citep{pei2020survey}. Our interval prediction therefore aligns with a realistic workflow in which products are first positioned into coarse price tiers (e.g., for rapid listing or benchmarking) and then refined into an exact quote when needed. Finally, given the highly right-skewed, long-tailed distribution of data-product prices, tier prediction provides a more robust and interpretable reference that is less sensitive to extreme values, while regression results offer finer granularity for applications requiring precise pricing.

\subsection{Feature representation}
\subsubsection{Text feature representation}
First, the product name, product detail, and product description are integrated to form the product information. Then, the following five text representation methods are applied for feature extraction.

\textbf{(1) Bag of words}

For the product information, we first remove stop words and carry out tokenization. Only English words with a minimum length of two characters are retained, and the top 500 most frequent terms across the corpus are selected. The detailed procedure is outlined below.

The input corpus is denoted as $\mathcal{D}=\{d_1,d_2,\dots,d_N\}$, where $d_{i}$ represents the $i\text{-}th$ document (i.e., product information). After tokenization and filtering, we construct a vocabulary $\mathcal{V}$ consisting of the top 500 most frequent words across the corpus. The Bag-of-Words (BOW) representation of the $i\text{-}th$ data product is then given by:
\begin{equation}
   \mathbf{x}_i = \phi(d_i), \quad
x_{ij} = \sum_{t \in T(d_i)} \mathbb{I}\{t = v_j\}, \quad j = 1,\ldots,|\mathcal{V}|,
\end{equation}
where $\phi(\cdot)$ denotes the feature mapping function, 
$\mathbf{x}_i = \phi(d_i)$ is the corresponding 
the BOW vector corresponding to document $d_{i}$. The element $x_{ij}$ represents the frequency of the  $j\text{-}th$ vocabulary term $v_j$ in document $d_i$. $T\left(d_i\right)$ is the multiset of tokens obtained after tokenizing document $d_{i}$, $\mathbb{I}\left\{t=v_j\right\}$ is an indicator function, which equals 1 if the token $t$ is equal to the vocabulary word $v_{j}$, and 0 otherwise.

\textbf{(2) TF-IDF}

Based on the Bag-of-Words (BOW) method, we compute the term frequency (TF). Specifically, for $i\text{-}th$ data product information $d_i$, the occurrence frequency of word $v_j$ is calculated as:
\begin{equation}
    \mathrm{tf}_{i j}=\frac{n_{i j}}{\sum_{k=1}^{|\mathcal{V}|} n_{i k}},
\end{equation}
\noindent where $n_{ij}$ denotes the number of times word $v_j$ appears in the data product information $d_i$.

We then compute the Inverse Document Frequency (IDF), which measures the discriminative power of a word $v_j$: 
\begin{equation}
    \operatorname{idf}_j=\log \frac{N}{1+\left|\left\{i: v_j \in d_i\right\}\right|},
\end{equation}
\noindent where $N$ denotes the total number of data products, and the denominator represents the number of documents containing the word $v_j$.

Next, we compute the TF-IDF weight for the $j$ -th word $v_j$ in the $i$ -th data product information $d_i$
\begin{equation}
    \mathrm{tfidf}_{i j}=\mathrm{tf}_{i j} \times \mathrm{idf}_j.
\end{equation}

Accordingly, the feature vector of the $i$ -th data product information is given by $\mathbf{x}_i=\left(\text { tfidf }_{i 1}, \text { tfidf }_{i 2}, \ldots, \text { tfidf }_{i|\mathcal{V}|}\right)$. Finally, we construct the document–term matrix: $X \in \mathbb{R}^{N \times|\mathcal{V}|}$, where each row corresponds to the TF-IDF representation of a data product information.

\textbf{(3) Word Embedding}

The matrices generated by BOW and TF-IDF are sparse and fail to capture semantic similarity between words. In contrast, Word2Vec produces dense vector representations that explicitly encode semantic relationships.
One common approach in Word2Vec is the Skip-gram model. Its objective is to learn word representations that can effectively predict the surrounding context within a sentence or document \citep{mikolovDistributedRepresentationsWords2013}.

Given a sequence of training words $x_1, x_2, x_3, \ldots, x_T$, the Skip-gram model aims to maximize the average log probability:
\begin{equation}
    \frac{1}{T} \sum_{t=1}^T \sum_{-c \leq j \leq c, j \neq 0} \log p\left(x_{t+j} \mid x_t\right),
\end{equation}
\noindent where $x_{t}$ denotes the center word and $x_{t+j}$ denotes a context word within a window of size $c=5$, with $j \in[-c, c] \backslash\{0\}$.

The conditional probability $p(x_{t+j} \mid x_t)$ is modeled using the softmax function:
\begin{equation}
    p(x_O \mid x_I)=\frac{\exp (v_{x_O}^{\prime}{ }^{\top} v_{x_I})}{\sum_{x=1}^\mathcal{V} \exp (v_x^{\prime}{ }^{\top} v_{x_I})},
\end{equation}
\noindent where $v_{x_I}$ and $v_x^{\prime}$ are the input (center word) and output (context word) vector representations of word $x$.

The trained word embedding matrix is denoted as $X \in \mathbb{R}^{|\mathcal{V}| \times d}$ where each word $x$ is represented by a vector $v(x)$. For a given document $d_{i}$, its set of word embeddings is $\left\{\mathbf{v}\left(x_{i 1}\right), \mathbf{v}\left(x_{i 2}\right), \ldots, \mathbf{v}\left(x_{i L_i}\right)\right\}$. The document vector is then constructed using average pooling:
\begin{equation}
    \mathbf{x}_i=\frac{1}{L_i} \sum_{t=1}^{L_i} \mathbf{v}\left(x_{i t}\right),
\end{equation}
\noindent where $L_{i}$ denotes the length of document $d_{i}$.

\textbf{(4) LDA model}

Given $N$ data product information documents, a vocabulary of size $T$, and $K$ latent topics, the modeling process of Latent Dirichlet Allocation (LDA) is as follows:

For each topic $k=1,\dots,K$
\begin{equation}
    \phi_k \sim \operatorname{Dirichlet}(\beta), \quad \phi_k \in \mathbb{R}^T, \sum_t \phi_{k, t}=1,
\end{equation}
\noindent where $\phi_k$ denotes the topic–word distribution, and $\beta$ is the Dirichlet prior for the topic–word distribution.

For each document $i=1,\dots,N$
\begin{equation}
    \theta_i \sim \operatorname{Dirichlet}(\alpha), \quad \theta_i \in \mathbb{R}^K, \sum_k \theta_{i, k}=1,
\end{equation}
\noindent where $\theta_i$ denotes the document–topic distribution, and $\alpha$ is the Dirichlet hyperparameter for the document–topic distribution.

For each position $n=1,\cdots,L_i$ in document $i$
\begin{equation}
    z_{i, n} \sim \operatorname{Categorical}\left(\theta_i\right), \quad x_{i, n} \sim \operatorname{Categorical}\left(\phi_{z_{i, n}}\right).
\end{equation}

The joint distribution of a single document is given by:
\begin{equation}
    p\left(\mathbf{x}_i, \mathbf{z}_i, \theta_i \mid \phi, \alpha\right)=p\left(\theta_i \mid \alpha\right) \prod_{n=1}^{L_i} p\left(z_{i, n} \mid \theta_i\right) p\left(x_{i, n} \mid z_{i, n}, \phi\right),
\end{equation}
\noindent where $\mathbf{x}_i=\left(x_{i, 1}, \ldots, x_{i, L_i}\right)$ represents all words in $i\text{-}th$ document,  $\mathbf{z}_i=\left(z_{i, 1}, \ldots, z_{i, L_i}\right)$ are the topic assignments of words, $p\left(\theta_i \mid \alpha\right)$ indicates that $\theta_i$ is drawn from the Dirichlet prior, $p\left(z_{i, n} \mid \theta_i\right)$ denotes sampling a topic for the $n\text{-}th$ word given $\theta_{i}$, and $p\left(x_{i, n} \mid z_{i, n}, \phi\right)$ denotes generating the word $x_{i,n}$ from the corresponding topic–word distribution.

The joint distribution of the entire corpus is:
\begin{equation}
\begin{aligned}
    p(\mathbf{X}, \mathbf{Z}, \Theta, \Phi \mid \alpha, \beta)&=\left(\prod_{k=1}^K p\left(\phi_k \mid \beta\right)\right)\\
    &\quad\times\left(\prod_{i=1}^N p\left(\theta_i \mid \alpha\right) \prod_{n=1}^{L_i} p\left(z_{i, n} \mid \theta_i\right) p\left(x_{i, n} \mid z_{i, n}, \Phi\right)\right).
    \end{aligned}
\end{equation}

By maximizing the likelihood, LDA ultimately estimates both the topic–word distributions and the document–topic distributions.

\textbf{(5) BERTopic}

LDA is based on the BOW representation which does not take contextual information into account. To address this limitation, \cite{grootendorstBERTopicNeuralTopic2022} proposed the BERTopic model. First, each document is transformed into a vector representation using the Sentence-BERT model. Next, the embedding vectors are reduced in dimensionality with UMAP (Uniform Manifold Approximation and Projection), followed by clustering with the HDBSCAN algorithm. The resulting clusters correspond to distinct topics. Subsequently, all documents within each cluster are merged into a single large document, and a class-based TF-IDF (c-TF-IDF) approach is introduced to extract the keyword distribution of each topic:
\begin{equation}
    X_{t, c}=t f_{t, c} \cdot \log \left(1+\frac{A}{t f_t}\right),
\end{equation}
\noindent where $t f_{t, c}$ denotes the frequency of term $t$ within topic $c$, $tf_t$ denotes the total frequency of term $t$ across all topics, and $A$ represents the average number of words per topic.

Based on BERTopic, one can derive both the most probable topic category for a given data product document and the corresponding probability distribution over topics.
\enlargethispage{2em}
\subsubsection{Numerical and categorical feature representation}
In addition to the data product information, several other relevant attributes are included, namely provider name, history, data sample, support Email, support URL, refund policy, and volume. 

\begin{itemize}
    \item The provider name is represented as a binary variable (0--1): if the provider is a publicly listed company, the value is set to 1; otherwise, it is 0. 
    \item History indicates whether the data product supports historical revision and future revision, each represented by a binary variable (0--1). 
    \item Data sample refers to whether the product provides sample data, also represented by a binary indicator. 
    \item Support email and support URL capture whether relevant contact information is available, again encoded as 0--1. 
    \item Refund policy describes the refund terms, categorized into five levels. For this feature, the classification was performed using DeepSeek, based on the textual descriptions of refund policies, with the detailed prompt instructions provided in the Appen-dix. 
    \item Volume refers to the number of datasets associated with a data product and is represented as an integer.
\end{itemize}

For data product industry, we employ DeepSeek (model version: DeepSeek-V3.2-Exp) to parse the product information and classify the products into 12 industries.\footnote{The industry classification of data products was conducted using the DeepSeek accessed via direct API calls. This choice was motivated by the reason that the classification task relies primarily on semantic understanding of textual descriptions rather than domain-specific labeled training data, which makes general-purpose large language model appropriate.} The industry taxonomy is based on Amazon’s organizational structure of data products, while making several domain-specific adjustments: for instance, agriculture and energy are separated as independent resource-based categories, and marketing data is further divided into e-commerce and offline retail scenarios. The specific industry classification scheme and corresponding prompt design are presented in the Appendix.

\subsection{Feature selection model}
To address the high-dimensional features that influence data product pricing, particularly semantic features, we employ the Maximum Relevance Minimum Redundancy (mRMR) method to select features that provide the highest information value for predicting data product prices \citep{gaoPredictingFinancialDistress2025}. The mRMR method is adopted because it effectively identifies a compact subset of features that maximizes prediction performance while reducing redundancy. The procedure of the mRMR method is described as follows.

The mRMR method is based on mutual information. Given two random variables $u$ and $v$ with probability distributions $p(u)$ and $p(v)$, respectively, their mutual information is defined as follows:
\begin{equation}
    I(u , v)=\sum_{u \in \Omega_u, v \in \Omega_v} p(u, v) \log \frac{p(u, v)}{p(u) p(v)},
\end{equation}
\noindent where $p(u,v)$ denotes the joint probability distribution of the two variables, and $\Omega_u$ and $\Omega_v$ represent the corresponding sample spaces of $u$ and $v$, respectively.

The goal of maximum relevance is to identify a feature subset $V$ containing $m$ features that maximizes its dependency on the data product price $p_c$. This is achieved by maximizing the relevance function $D(V,p_c)$, defined as:
\begin{equation}
    \max D(V, p_c), D=\frac{1}{|V|} \sum_{u_i \in S} I(u_i, p_c),
\end{equation}
\noindent where $I(u_i , p_c)$ denotes the mutual information between an individual feature $u_i$ and the target variable $p_c$.

However, the features selected solely based on maximum relevance may still exhibit high redundancy. Minimum redundancy aims to eliminate redundant features without deteriorating the model’s predictive performance. The redundancy measure under this criterion is defined as:
\begin{equation}
    \min R(V), R=\frac{1}{|V|^2} \sum_{u_i ; u_j \in V} I(u_i, u_j).
\end{equation}

The mRMR method combines the maximum relevance and minimum redundancy criteria by defining an operator $\Phi(D, R)$ that jointly accounts for both measures. The optimal feature subset is then obtained by maximizing this objective function:
\begin{equation}
    \max \Phi(D, R), \Phi=D-R.
\end{equation}

We employ an incremental search strategy to identify a near-optimal feature subset according to the above criteria. The selection process is defined as follows:
\begin{equation}
    \max_{u_j \in X-V_{m-1}}[I(u_j , c)-\frac{1}{m-1} \sum_{u_i \in V_{m-1}} I(u_j : u_i)],
\end{equation}
\noindent where the $m$-th feature is selected from the set $X-V_{m-1}$ so as to maximize the objective function $\Phi(\cdot)$.

Based on the mRMR method, we are able to select features that exhibit strong relevance to the data product price while maintaining low redundancy among the selected variables.

\subsection{Data price prediction model}
In this study, there are two types of prediction tasks: a continuous data product price regression task and a price-range classification task. The price-range classification is formulated as a multi-class problem. For the multi-class problem, we employ the one-vs-rest strategy to decompose the multi-class setting into a series of binary classification tasks. Therefore, the subsequent discussion of classification methods mainly focuses on the binary-classification perspective. The models employed in this study include Logistic Regression (LR), Artificial Neural Networks (ANN), Decision Trees (DT), Support Vector Machines (SVM), Random Forests (RF), and XGBoost.
\subsubsection{LR model}
Given a numerical feature vector and a textual semantic vector $\boldsymbol{x}=\left(x_1, x_2, \ldots, x_d\right)$, the linear model is defined as: 
\begin{equation}
    y=w_1 x_1+w_2 x_2+\ldots+w_d x_d+b=\boldsymbol{w}^T \boldsymbol{x}+b,
\end{equation}
\noindent where $\boldsymbol{w}=\left(w_1, w_2, \ldots, w_d\right)$ denotes the weight vector and $b$ is the bias term. In the regression task, $y$ represents the predicted price of a data product.

For the data-product price classification task, the linear score $z=\boldsymbol{w}^{\mathrm{T}} \boldsymbol{x}+b$ is mapped to a binary outcome (0/1). Using the logistic function, the model output $y$ is defined as:
\begin{equation}
    y=\frac{1}{1+e^{-z}}=\frac{1}{1+e^{-\left(\boldsymbol{w}^T \boldsymbol{x}+b\right)}}.
\end{equation}

Applying the logit transformation yields:
\begin{equation}
    \ln \frac{y}{1-y}=\boldsymbol{w}^T \boldsymbol{x}+b.
\end{equation}

If $y$ is interpreted as the posterior probability $p(y=1 \mid \boldsymbol{x})$, the expression becomes:
\begin{equation}
    \ln \frac{p(y=1 \mid \boldsymbol{x})}{p(y=0 \mid \boldsymbol{x})}=\boldsymbol{w}^T \boldsymbol{x}+b.
\end{equation}

The parameters $\boldsymbol{w}$ and $b$ are estimated via maximum likelihood. Based on the estimated parameters, the LR model produces the final classification predictions.
\subsubsection{ANN model}
Given a training dataset $\left\{\left(\boldsymbol{x}_1, y_1\right),\left(\boldsymbol{x}_2, y_2\right), \ldots,\left(\boldsymbol{x}_d, y_d\right)\right\}$, the first layer of the ANN is the input layer, containing $m$ neurons $\boldsymbol{x}=\left(x_1, x_2, \ldots, x_m\right)$. The designed ANN consists of multiple hidden layers. Taking a single hidden layer as an illustrative example. For the first hidden layer, the output of neuron $j$ is computed as:
\begin{equation}
    o_j^{(1)}=\theta\left(\sum_{i=1}^d \boldsymbol{w}_{i, j}^{(1)} \boldsymbol{x}_i+b_j\right),
\end{equation}
\noindent where $\theta(\cdot)$ denotes the activation function, $\boldsymbol{w}_{i,j}$ represents the connection weight between input neuron $i$ and hidden neuron $j$, and $b_j$ is the bias term.

The outputs of neurons in all subsequent hidden layers are computed using the similar computation. The final layer is the output layer, consisting of multiple neurons. Based on the output layer, the network directly outputs the predicted price for regression tasks, while for classification tasks it assigns the label corresponding to the class with the highest predicted probability. For classification task, this layer applies the softmax activation function, and the probability of class $i$ is given by:
\begin{equation}
    p_i=\frac{e^{z_i}}{\sum_{j=1}^n e^{z_j}}.    
\end{equation}
\subsubsection{DT model}
A decision tree model consists of a root node, several internal nodes, and leaf nodes, where the leaf nodes represent the final decisions and all other nodes correspond to attribute tests. The tree is constructed by recursively partitioning the feature space to form a sequence of tests from the root to the leaves. In this study, we employ the Classification and Regression Trees (CART) model for both regression and classification tasks: the regression trees use squared error as the splitting criterion, while the classification trees adopt the gini index.
\subsubsection{SVM model}
Given a training dataset $\left\{\left(\boldsymbol{x}_1, y_1\right),\left(\boldsymbol{x}_2, y_2\right), \ldots,\left(\boldsymbol{x}_d, y_d\right)\right\}$, suppose that no separating hyperplane exists in the original feature space that can perfectly divide the two classes. Let $\phi(\boldsymbol{x})$ denote the mapping of $\boldsymbol{x}$ into a higher-dimensional feature space. A hyperplane in this feature space is expressed as:
\begin{equation}
    f(\boldsymbol{x})=\boldsymbol{w}^T \phi(\boldsymbol{x})+b,
\end{equation}
\noindent where $\boldsymbol{w}$ and $b$ represent the weight vector and bias term, respectively. To determine the separating hyperplane with the maximum margin, the model must satisfy the following optimization problem:
\begin{equation}
\begin{aligned}
\min_{\boldsymbol{w}, b} \quad & \frac{1}{2}\|\boldsymbol{w}\|^2 \\
\text{s.t.} \quad & y_i(\boldsymbol{w}^T \phi(\boldsymbol{x}_i)+b) \ge 1,\quad i=1,\ldots,n.
\end{aligned}
\end{equation}

Using the Lagrange multiplier method and introducing soft margins, the dual optimization problem can be formulated as:
\begin{equation}
\begin{aligned}
    \max \quad & _\alpha \sum_{i=1}^n \alpha_i-\frac{1}{2} \sum_{i=1}^n \sum_{j=1}^n \alpha_i \alpha_j y_i y_j \phi\left(\boldsymbol{x}_i\right)^T \phi\left(\boldsymbol{x}_j\right) \\
    \text { s.t. } \quad & \sum_{i=1}^n \alpha_i y_i=0 \\ 
    \quad & 0 \leq \alpha_i \leq C, i=1,2, \ldots, \mathrm{n},
\end{aligned}
\end{equation}
\noindent where $\alpha_i$ are the Lagrange multipliers and $C$ is the regularization constant.

Because the dimensionality of the feature space may be high, directly computing $\phi\left(\boldsymbol{x}_i\right)^T \phi\left(\boldsymbol{x}_j\right)$ may be computationally expensive. Therefore, SVMs employ kernel functions defined as:
\begin{equation}
    k\left(x_i, x_j\right)=\phi\left(\boldsymbol{x}_i\right)^T \phi\left(\boldsymbol{x}_j\right),
\end{equation}
\noindent with common choices including the linear, polynomial, radial basis function (RBF), sigmoid, and Laplacian kernels.

When the solution satisfies the Karush–Kuhn–Tucker (KKT) conditions, the solutions to both the primal and dual problems can be obtained. Based on the learned parameters $\boldsymbol{w}$ and $b$, the SVM model produces its predictions.

The above formulation applies to classification SVMs. For regression tasks, the Support Vector Regression (SVR) optimization problem is expressed as:
\begin{equation}
    \min _{\boldsymbol{w}, b} \frac{1}{2}\|\boldsymbol{w}\|^2+C \sum_{i=1}^m \ell_c\left(f\left(\boldsymbol{x}_i\right)-y_i\right),
\end{equation}
\noindent where $C$ is the regularization constant and $\ell_c$ is $\epsilon$-insensitive loss function defined as:
\begin{equation}
    \ell_\epsilon(z)= \begin{cases}0, & \text { if }|z| \leqslant \epsilon  \\ |z|-\epsilon, & \text { otherwise } \end{cases}.
\end{equation}
\subsubsection{RF model}
Since a decision tree is trained using all available samples and all features, it is prone to overfitting. To mitigate this risk, the Random Forest (RF) model constructs an ensemble of decision trees generated through randomization. Given a dataset with $n$ samples, each containing $K$ features, each decision tree in the random forest is built as follows: a bootstrap sample of $m$ instances is drawn from the original dataset with replacement ($m < n$), and a random subset of $k$ features is selected ($k<K$). A single decision tree is then trained using this subset of samples and features. By repeatedly applying this random sampling process, multiple diverse trees are produced. The combination of sample-level randomness and feature-level randomness ensures sufficient diversity across the trees in the ensemble.
\subsubsection{XGBoost model}
XGBoost is an ensemble algorithm based on Gradient Tree Boosting that can be applied to both classification and regression tasks. Given a training dataset $\{(\boldsymbol{x}_1, y_1),(\boldsymbol{x}_2, y_2), \ldots,(\boldsymbol{x}_d, y_d)\}$, suppose the XGBoost model consists of $m$ decision trees. The prediction for the $i$-th sample is then computed as: 
\begin{equation}
    \hat{y}_i=F_m\left(x_i\right)=F_{m-1}\left(x_i\right)+f_m\left(x_i\right),
\end{equation}
\noindent where $f_m(\cdot)$ denotes the $m$-th decision tree. Each tree maps a sample to one of its leaf nodes based on the feature splits. Every leaf node carries a weight score $\omega$, which serves as the tree’s prediction for any sample falling into that leaf.

The final prediction of the model is obtained by summing the leaf scores $\omega$ across all trees. The objective function of XGBoost is defined as:
\begin{equation}
    Loss=\sum_{i=1}^n L\left(y_i, \hat{y}_i\right)+\sum_{k=1}^m \Omega\left(f_k\right),
\end{equation}
\noindent where the first term measures the loss or error between the predicted and true values, and the second term is a regularization component that controls model complexity. The regularization term encourages simpler tree structures and helps reduce the risk of overfitting.

\subsection{Performance evaluation}
The dataset is divided using 5-fold cross-validation. For the regression task, the evaluation metrics included Mean Squared Error (MSE), the coefficient of determination ($R^2$) and Mean Absolute Percentage Error (MAPE). The calculation formulas are as follows:
\begin{equation}
    MSE=\frac{1}{n}\sum_{i=1}^{n}{(y_i-{\hat{y}}_i)}^2,
\end{equation}
\begin{equation}
R^2=1-
\frac{\sum_{i=1}^{n} \left( y_i - \hat{y}_i \right)^2}
{\sum_{i=1}^{n} \left( y_i - \bar{y} \right)^2},
\end{equation}
\begin{equation}
    MAPE=\frac{1}{n} \sum_{i=1}^{n}\frac{\left|y_i-{\hat{y}}_i\right|}{y_i},
\end{equation}
\noindent where ${\hat{y}}_i$ denotes the predicted value, $y_i$ represents the actual value and $\bar{y} = \frac{1}{n} \sum_{i=1}^{n} y_i$.

For the classification task, the evaluation metrics include Accuracy, AUC, and F1-Score. Accuracy is defined as the proportion of correctly classified samples relative to the total number of samples. Both AUC and F1-Score are computed as the average of the corresponding metrics obtained from multiple binary classification subproblems under the one-vs-rest strategy. In binary classification, AUC represents the area under the ROC curve, while the F1-Score is the harmonic-weighted combination of Precision and Recall. These three metrics are used to comprehensively evaluate the performance of the classification models.
\newpage
\section{Results} \label{sec:result}

\subsection{Data description}
\subsubsection{Sample}
In this study, we collect data from Amazon Data Exchange, which is a comprehensive marketplace for third-party datasets. It provides a secure and compliant environment, seamlessly integrates with Amazon Web Services and third-party tools and services, and supports unified billing and subscription management. We exclude data products with free prices, and primarily focuses on subscription-based data products. As a result, the sample consists of 2,073 data products.

\subsubsection{Feature}
The features of the dataset include product name, product detail, product description, provider name, volume, history, data sample, support email, support URL, and refund policy, as summarized in Table \ref{table:feature description}. The descriptive statistics of the features are presented in Table \ref{table: descriptive statistics}. 
We can see from Table \ref{table: descriptive statistics} that in term of the feature (Data provider is listed), the mean value is only 0.08, indicating that the majority of data providers are non-listed enterprises or institutions, with only a small proportion being publicly listed companies. The Data volume number has a mean of 1.74, a standard deviation of 4.95, and a maximum value of 89, suggesting substantial variation in dataset size across different data products. For Sensitive level, the mean value is 0.25, implying that most data products do not contain sensitive information. Regarding version-related variables, the mean value of Future version is 0.94, indicating that most products support future updates, while the mean of Historical version is 1.68, showing that a considerable number of products provide access to historical data.

\begin{table}
    \centering
    \caption{Feature description}
    \label{table:feature description}
    % \fontsize{11pt}\selectfont
    % \resizebox{\textwidth}{!}{
    \small
    \begin{tabular}{p{3cm}p{6cm}l}
    \hline
        Feature name & Feature description & Value range \\
    \hline
        Product name & Title of the data product  & Text\\
        Product detail & A short summary describing the core content of the dataset & Text  \\
        Product description & A detailed explanation of the dataset & Text \\
    \hline
        Listed Provider & Whether the data provider is a publicly listed company & \{0,1\}\\
        % Provider link & URL & Dummy variable\\
    \hline
        % Product usage information & How the data product can be accessed, queried, and used & string \\
        Volume & The size or quantity of the dataset & Numerical \\
        Historical version & The temporal coverage of the dataset, including access to past revisions 
        & \{0,1,2\}
        \\
        Future version & The temporal coverage of the dataset, including access to future updates & \{0,1\} \\
        
        Sensitive & Whether the dataset contains sensitive information & \{0,1,2\} \\
        % Data table & Metadata describing the tables contained within the data product & \\
    \hline
        Data sample & A link or reference to a publicly accessible sample of the dataset & \{0,1\} \\
    
        Support contact email address & Contact for technical or product-related queries & \{0,1\} \\
        
        Support contact URL & Support website or portal 
        & \{0,1\}
        \\
        Refund policy & Rules governing refund eligibility for the data subscription & \{0,1,2,3,4\} \\
    \hline
    \end{tabular}
    % }
\end{table}

\begin{table}[h]
    \centering
    \caption{Descriptive statistics of data product features}
    \label{table: descriptive statistics}
    \small
    \begin{tabular}{lllllll}
    \hline
        Feature & Average & Std & Max & Min & Skewness & Kurtosis  \\ 
    \hline
        Data provider is listed & 0.08  & 0.27  & 1.00  & 0.00  & 3.16  & 7.96   \\ 
        Data volume number & 1.74  & 4.95  & 89.00  & 1.00  & 12.65  & 187.76   \\
        Sensitive level & 0.25  & 0.65  & 2.00  & 0.00  & 2.26  & 3.22   \\ 
        Future version & 0.94  & 0.25  & 1.00  & 0.00  & -3.56  & 10.66   \\ 
        Historical version & 1.68  & 0.56  & 2.00  & 0.00  & -1.56  & 1.46   \\ 
        Data sample & 1.00  & 0.05  & 1.00  & 0.00  & -20.29  & 409.60   \\ 
        Email & 1.00  & 0.05  & 1.00  & 0.00  & -20.29  & 409.60   \\ 
        URL & 1.00  & 0.05  & 1.00  & 0.00  & -20.29  & 409.60   \\
        Refund & 0.65  & 1.12  & 4.00  & 0.00  & 1.46  & 0.69   \\ 
        \hline
    \end{tabular}
\end{table}

In terms of product support features, the variables Data sample, Email, and URL all have mean values of 1 and very small standard deviations (0.05), implying that almost all products offer sample data, technical support via email, and an accessible support website link. As for the Refund policy, the mean value is 0.65, indicating that most products include some form of refund provision. However, the standard deviation of 1.12 suggests a high degree of variability, reflecting multiple levels of refund rules across different products.

The distribution of different industries is presented in Table \ref{table:distribution of different industry}. To evaluate the reliability of the classification, we conducted a manual validation exercise. Specifically, we drawed a random sample of 120 data products from the full dataset, covering all 12 industry categories, with 10 observations per category. Then we reviewed and assigned each sample to an industry label based on product’s textual title and description. The manually assigned labels were then compared with the generated labels to assess accuracy. Validation results indicate a high level of classification accuracy of 0.94 (113 correct classifications out of 120). These results suggest that misclassifications are limited and do not materially affect the overall industry distribution patterns. Manual validation results of industry classification accuracy are reported in Table \ref{table:Manual validation results of industry classification accuracy}.

\begin{table}
    \centering
    \caption{Distribution of different industries}
    \label{table:distribution of different industry}
    \begin{tabular}{lll}
    \hline
        Industry & Count & Percentage (\%) \\ 
    \hline
        E-commerce and Business & 548 & 16.23 \\
        Financial Services & 469  & 13.89 \\ 
        Retail and Location & 459 & 13.60 \\
        Public Sector & 300  & 8.89 \\ 
        Healthcare and Life Sciences& 242 & 7.17  \\ 
        Media and Entertainment& 234 & 6.93 \\ 
        Cars and Automotive & 225  & 6.66 \\ 
        Environmental& 208 & 6.16\\ 
        Gaming& 201  & 5.95 \\
        Telecommunications& 183 & 5.42 \\ 
        Resources & 160 & 4.74\\
        Manufacturing & 147 & 4.35\\
        Total & 3376 & 100 \\
    \hline
    \end{tabular}
\end{table}

\begin{table}
    \centering
    \caption{Manual validation results of industry classification accuracy}
    \label{table:Manual validation results of industry classification accuracy}
    \begin{tabular}{llll}
        \hline
        Industry & Correct & Total & Accuracy  \\
        \hline
        E-commerce and Business Data & 10 & 10 & 1.00  \\ 
        Financial Services & 10 & 10 & 1.00  \\ 
        Retail and Location Data & 10 & 10 & 1.00  \\ 
        Public Sector Data & 9 & 10 & 1.00  \\ 
        Healthcare and Life Sciences Data & 10 & 10 & 1.00  \\ 
        Media and Entertainment Data & 10 & 10 & 1.00  \\ 
        Cars and Automotive Data & 9 & 10 & 0.90  \\ 
        Environmental Data & 9 & 10 & 0.90  \\ 
        Gaming Data & 8 & 10 & 0.80  \\ 
        Telecommunications Data & 10 & 10 & 1.00  \\ 
        Resources Data & 9 & 10 & 0.90  \\ 
        Manufacturing Data & 9 & 10 & 0.90  \\ 
        Total & 113 & 120 & 0.94  \\
        \hline
    \end{tabular}
\end{table}

Overall, data products are mainly concentrated in three primary categories: E-comme-rce and Business Data, Financial Services, and Retail and Location Data, with counts of 548, 469, and 459 respectively, representing the largest proportions in the dataset. Following these are Public Sector Data (300) and Healthcare and Life Sciences Data (242), indicating notable representation in government and medical domains. Media and Entertainment Data (234) and Cars and Automotive Data (225) fall into the medium range, reflecting moderate activity in these sectors. In contrast, Environmental Data (208), Gaming Data (201), Telecommunications Data (183), Resources Data (160), and Manufacturing Data (147) account for smaller shares. It is worth noting that a single data product may be classified into multiple industries. Therefore, the total number reported in Table \ref{table:distribution of different industry} exceeds the number of data products in the sample.

\subsubsection{Data product price}
\begin{figure}
    \centering
    \includegraphics[width=0.9\linewidth]{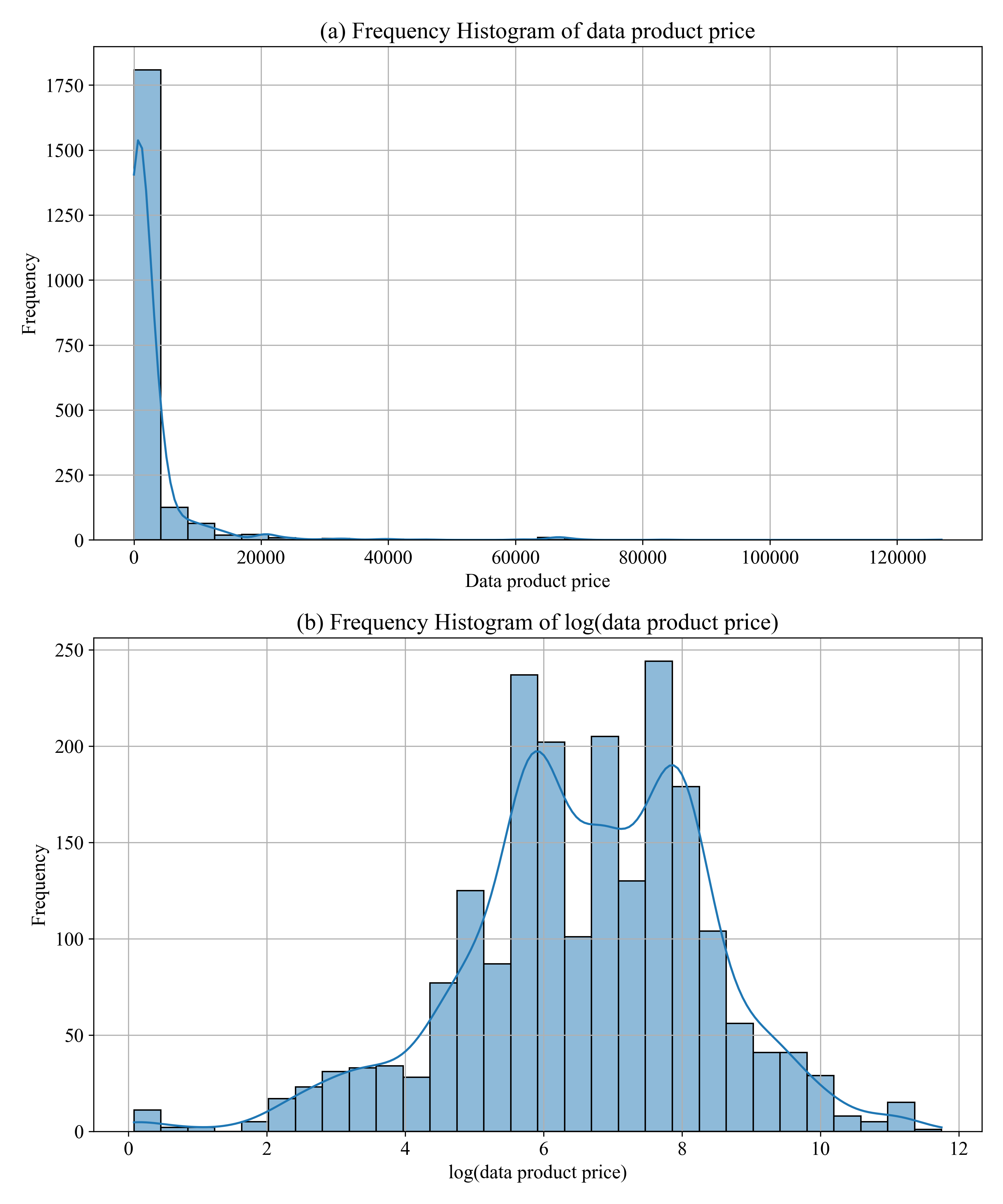}
    \caption{Distribution analysis of data product price}
    \label{fig: distribution of data price}
\end{figure}

Figure \ref{fig: distribution of data price} illustrates the frequency distribution of data product prices and their logarithmic transformations.
First, as shown in the original price distribution chart (a) in Figure \ref{fig: distribution of data price}, data product prices exhibit a highly right-skewed distribution, with the majority of products concentrated in the low-price range (approximately 0--3,000 USD). In contrast, only a few high-priced products exist, yet their price range is extremely wide, reaching up to over 100,000 USD. This pattern indicates a pronounced long-tail effect in the data product market. Secondly, the logarithmic price distribution chart (b) shows that after the log transformation, the distribution becomes notably more symmetric and approximately normal, suggesting that the transformation effectively reduces the influence of extreme values. The log (price) values are primarily concentrated between 5 and 8, corresponding to an original price range of roughly 150--3,000 USD, implying that most data products in the market are priced within this range.

In addition to predicting the exact prices of data products, we also conduct price interval prediction. Specifically, the prices are divided into four quartiles, resulting in five distinct ranges from high to low. 
After categorizing the prices, the corresponding price ranges are defined as follows:  Class 0: $\leq$ 208.33; Class 1: 212.5-416.67; Class 2: 418.5-1,250.0; Class 3: 1,258.33-3,175.0; Class 4: $\geq$ 3,200.0.
Such interval-based prediction can provide data providers and data consumers with an approximate price range for reference. 

\subsection{Regression results comparison based on different feature representation methods}
To compare the predictive performance of different text feature representation methods (BOW, TF-IDF, Word2Vec, LDA, and BERTopic) on data-product pricing, multiple prediction models are employed, including Linear Regression (LR), Artificial Neural Network (ANN), Decision Tree Regression (DT), Support Vector Regression (SVR), Random Forest (RF), and XGBoost. The prediction results are shown in Table \ref{table:table-different feature representation} where MV represents the mean values across the six predictive models, Rank indicates the ranking results based on the MV value, and the values in bold font indicate the best results among different text representation methods under the same prediction model.
\begin{table}
    \centering
    \caption{Regression prediction results of different text feature representation methods. All results are reported as the mean and standard deviation over five-fold cross-validation.}
    \label{table:table-different feature representation}
    \resizebox{\textwidth}{!}{
    \begin{tabular}{lcccccccc}
    \hline
        \textbf{Method} & \textbf{LR} & \textbf{ANN} & \textbf{DT} & \textbf{SVR} & \textbf{RF} & \textbf{XGBoost} & \textbf{MV} & \textbf{Rank } \\ \hline
        ~ & \multicolumn{8}{c}{MSE}\\ 
        \cline{2-9}
        BOW & 6.7360 & 1.3428 & \textbf{1.6485} & 1.5015 & \textbf{0.8810} & 1.2022 & 2.2187 & 4  \\
        ~ & ($\pm$4.2524) & ($\pm$0.168) & ($\pm$0.1494) & ($\pm$0.1251) & ($\pm$0.1350) & ($\pm$0.1211)  & ($\pm$2.2287) & ~ \\ 
        
        TFIDF & 1.8123 & \textbf{1.3382} & 1.9973 & \textbf{1.4823} & 1.0389 & 1.2597 & 1.4881 & 2  \\
        ~ & ($\pm$0.2411) & ($\pm$0.1462) & ($\pm$0.2758) & ($\pm$0.1474) & ($\pm$0.1256) & ($\pm$0.1559)  & ($\pm$0.3580) & ~ \\
        
        Word2Vec & \textbf{1.5292} & 1.3818 & 2.1240 & 1.4928 & 1.0396 & \textbf{1.1631} & \textbf{1.4551} & 1  \\
        ~ & ($\pm$0.2032) & ($\pm$0.2341) & ($\pm$0.2341) & ($\pm$0.1354) & ($\pm$0.1262) & ($\pm$0.1393)  & ($\pm$0.3788) & ~ \\
        
        LDA & 2.2387 & 1.7753 & 2.2764 & 2.0380 & 1.2628 & 1.5415 & 1.8555 & 3  \\
        ~ & ($\pm$0.1066) & ($\pm$0.1568) & ($\pm$0.2848) & ($\pm$0.1304) & ($\pm$0.117) & ($\pm$0.1121)  & ($\pm$0.4034) & ~ \\
        
        BERTopic & 2.7145 & 2.2943 & 2.9253 & 2.4133 & 1.6244 & 1.9376 & 2.3182 & 5  \\
        ~ & ($\pm$0.0871) & ($\pm$0.2036) & ($\pm$0.2367) & ($\pm$0.0777) & ($\pm$0.1539) & ($\pm$0.1196) & ($\pm$0.4820) & ~ \\
        
        \hline

        ~ & \multicolumn{8}{c}{$R^2$}\\ 
        \cline{2-9}
        BOW & -1.1489 & 0.5703 & \textbf{0.4718} & 0.5204 & \textbf{0.7182} & 0.6157 & 0.2913  & 4  \\
        ~ & ($\pm$1.3601) & ($\pm$0.0568) & ($\pm$0.0581) & ($\pm$0.0314) & ($\pm$0.044) & ($\pm$0.0371)  & ($\pm$0.7106) & ~ \\
        
        TFIDF & 0.4221 & \textbf{0.5725} & 0.3614 & \textbf{0.5270} & 0.6684 & 0.5977 & 0.5249  & 2  \\
        ~ & ($\pm$0.061) & ($\pm$0.0424) & ($\pm$0.0887) & ($\pm$0.0331) & ($\pm$0.0341) & ($\pm$0.0446)  & ($\pm$0.1144) & ~ \\
        
        Word2Vec & \textbf{0.5106} & 0.5569 & 0.3202 & 0.5231 & 0.6679 & \textbf{0.6286} & \textbf{0.5346}  & 1  \\
        ~ & ($\pm$0.0682) & ($\pm$0.0828) & ($\pm$0.0806) & ($\pm$0.0369) & ($\pm$0.0368) & ($\pm$0.0386)  & ($\pm$0.1215) & ~ \\
        
        LDA & 0.2845 & 0.4321 & 0.2704 & 0.3487 & 0.5962 & 0.5073 & 0.4065  & 3  \\ 
        ~ & ($\pm$0.0212) & ($\pm$0.0522) & ($\pm$0.1044) & ($\pm$0.033) & ($\pm$0.0373) & ($\pm$0.0316)  & ($\pm$0.1293) & ~ \\
        
        BERTopic & 0.1323 & 0.2673 & 0.0654 & 0.2284 & 0.4813 & 0.3811 & 0.2593  & 5 \\
        ~ & ($\pm$0.0102) & ($\pm$0.0519) & ($\pm$0.0629) & ($\pm$0.0175) & ($\pm$0.0391) & ($\pm$0.0231) & ($\pm$0.1542) & ~ \\
        \hline

        ~ & \multicolumn{8}{c}{MAPE} \\ \cline{2-9}
        BOW & 0.3256 & \textbf{0.1520} & \textbf{0.1387} & 0.4814 & \textbf{0.1647} & 0.2859 & 0.2581  & 2  \\
        ~ & ($\pm$0.0651) & ($\pm$0.0229) & ($\pm$0.0142) & ($\pm$0.104) & ($\pm$0.0298) & ($\pm$0.0597)  & ($\pm$0.1337) & ~ \\ 
        
        TFIDF & 0.3068 & 0.1998 & 0.1435 & 0.4432 & 0.2067 & 0.2425 & \textbf{0.2571}  & 1  \\
        ~ & ($\pm$0.0542) & ($\pm$0.0571) & ($\pm$0.022) & ($\pm$0.0978) & ($\pm$0.0478) & ($\pm$0.0379)  & ($\pm$0.1059) & ~ \\
        
        Word2Vec & \textbf{0.2992} & 0.2197 & 0.1855 & \textbf{0.4221} & 0.2119 & \textbf{0.2247} & 0.2605  & 3  \\
        ~ & ($\pm$0.0802) & ($\pm$0.0373) & ($\pm$0.0807) & ($\pm$0.0906) & ($\pm$0.0641) & ($\pm$0.062)  & ($\pm$0.0878) & ~ \\
        
        LDA & 0.4721 & 0.2286 & 0.2117 & 0.5085 & 0.2287 & 0.3810 & 0.3384  & 4  \\
        ~ & ($\pm$0.088) & ($\pm$0.0414) & ($\pm$0.0727) & ($\pm$0.1028) & ($\pm$0.0521) & ($\pm$0.0996)  & ($\pm$0.1332) & ~ \\
        
        BERTopic & 0.5720 & 0.2714 & 0.2777 & 0.5550 & 0.2200 & 0.3795 & 0.3793  & 5 \\
        ~ & ($\pm$0.1096) & ($\pm$0.0528) & ($\pm$0.1112) & ($\pm$0.1088) & ($\pm$0.0481) & ($\pm$0.0763) & ($\pm$0.1519) & ~ \\
        \hline
    \end{tabular}
    }
\end{table}

Based on the results in Table \ref{table:table-different feature representation}, we first find that different text feature representation methods exhibit varying predictive performances. Word2Vec achieves the best overall results, ranking first in both MSE and $R^2$, indicating that its semantic representation provides the highest predictive capability for data-product pricing. TF-IDF follows closely, ranking first in MAPE, which suggests that traditional statistical methods remain robust when dealing with high-dimensional and sparse text data. In contrast, BERTopic performs the worst, likely due to insufficient data volume or overly coarse clustering granularity, leading to poor generalization of its topic representations. 
Second, regarding the results of different predictive models: LR performs relatively poorly across all feature representation methods, while ensemble models such as RF and XGBoost achieve better results. This indicates that the relationship between data-product prices and their textual features is inherently nonlinear.

% \subsection{Prediction results based on mRMR method}
\subsection{Classification results based on different feature representation methods}

In the classification task, we compare the predictive capabilities of different text representation methods (BOW, TF-IDF, Word2Vec, LDA, and BERTopic) in forecasting data product prices. Several prediction models are employed for this experiment, including LR, ANN, DT, SVM, RF, and XGBoost. Table \ref{table:classifiction task} presents the classification prediction results obtained using different text representation methods. In the Table, MR represents the mean results across the six predictive models, Rank indicates the ranking results based on the MR values, and the values in bold font indicate the best results among different text representation methods under the same prediction model.
\begin{table}
    \centering
    \caption{Classification prediction results of different text feature representation methods. All results are reported as the mean and standard deviation over five-fold cross-validation.}
    \label{table:classifiction task}
    \resizebox{\textwidth}{!}{
    \begin{tabular}{lcccccccc}
    \hline
        Method & LR & ANN & DT & SVM & RF & XGBoost & MV & Rank  \\ \hline
        ~ & \multicolumn{8}{c}{Accuracy} \\ 
        \cline{2-9}
        BOW & \textbf{0.7125} & \textbf{0.7279} & \textbf{0.7164} & \textbf{0.6927} & 0.7530 & 0.7636 & \textbf{0.7277}  & 1  \\
        ~ & ($\pm$0.02) & ($\pm$0.0196) & ($\pm$0.0176) & ($\pm$0.0177) & ($\pm$0.0293) & ($\pm$0.0246)  & ($\pm$0.0265) & ~ \\
        
        TFIDF & 0.6802 & 0.7217 & 0.6980 & 0.6884 & 0.7472 & 0.7453 & 0.7135  & 2  \\
        ~ & ($\pm$0.0236) & ($\pm$0.0225) & ($\pm$0.014) & ($\pm$0.0218) & ($\pm$0.0375) & ($\pm$0.0254)  & ($\pm$0.029) & ~ \\
        
        Word2Vec & 0.6701 & 0.6922 & 0.6773 & 0.6691 & \textbf{0.7689} & \textbf{0.7685} & 0.7077  & 3  \\
        ~ & ($\pm$0.0231) & ($\pm$0.0154) & ($\pm$0.0176) & ($\pm$0.0169) & ($\pm$0.0262) & ($\pm$0.024)  & ($\pm$0.048) & ~ \\ 
        
        LDA & 0.5552 & 0.6459 & 0.6474 & 0.6045 & 0.6922 & 0.6927 & 0.6397  & 4  \\
        ~ & ($\pm$0.0161) & ($\pm$0.0297) & ($\pm$0.0305) & ($\pm$0.0188) & ($\pm$0.0293) & ($\pm$0.0157)  & ($\pm$0.053) & ~ \\ 
        
        BERTopic & 0.4462 & 0.5962 & 0.5986 & 0.5080 & 0.6044 & 0.6300 & 0.5639  & 5 \\
        ~ & ($\pm$0.0195) & ($\pm$0.0227) & ($\pm$0.0135) & ($\pm$0.0142) & ($\pm$0.014) & ($\pm$0.0235) & ($\pm$0.071) & ~ \\
        
        \hline
        
        ~ & \multicolumn{8}{c}{AUC}  \\ 
        \cline{2-9}
        BOW & \textbf{0.9303} & \textbf{0.9282} & \textbf{0.8265} & \textbf{0.9178} & 0.9454 & \textbf{0.9486} & \textbf{0.9161}  & 1  \\
        ~ & ($\pm$0.0069) & ($\pm$0.0092) & ($\pm$0.0079) & ($\pm$0.0095) & ($\pm$0.0178) & ($\pm$0.0107)  & ($\pm$0.0454) & ~ \\
        
        TFIDF & 0.9145 & 0.9188 & 0.8222 & 0.9117 & 0.9392 & 0.9445 & 0.9085 & 2  \\
        ~ & ($\pm$0.0113) & ($\pm$0.0072) & ($\pm$0.0101) & ($\pm$0.0107) & ($\pm$0.024) & ($\pm$0.0123)  & ($\pm$0.0444) & ~ \\
        
        Word2Vec & 0.9011 & 0.9183 & 0.7987 & 0.8980 & \textbf{0.9480} & 0.9462 & 0.9017  & 3  \\
        ~ & ($\pm$0.0089) & ($\pm$0.0067) & ($\pm$0.0073) & ($\pm$0.0091) & ($\pm$0.0164) & ($\pm$0.0093)  & ($\pm$0.0548) & ~ \\
        
        LDA & 0.8474 & 0.8979 & 0.7839 & 0.8605 & 0.9182 & 0.9164 & 0.8707  & 4  \\
        ~ & ($\pm$0.0075) & ($\pm$0.0129) & ($\pm$0.0145) & ($\pm$0.0064) & ($\pm$0.0191) & ($\pm$0.0062)  & ($\pm$0.0515) & ~ \\
        
        BERTopic & 0.7418 & 0.8531 & 0.7707 & 0.7956 & 0.8569 & 0.8814 & 0.8166  & 5  \\
        ~ & ($\pm$0.0101) & ($\pm$0.0106) & ($\pm$0.0076) & ($\pm$0.0096) & ($\pm$0.0094) & ($\pm$0.0121) & ($\pm$0.0553) & ~ \\
        \hline
        
        ~ & \multicolumn{8}{c}{F1-Score} \\ 
        \cline{2-9}
        BOW & \textbf{0.7132} & \textbf{0.7285} & \textbf{0.7167} & \textbf{0.6926} & 0.7535 & 0.7641 & \textbf{0.7281}  & 1  \\
        ~ & ($\pm$0.0198) & ($\pm$0.02) & ($\pm$0.0174) & ($\pm$0.0172) & ($\pm$0.0303) & ($\pm$0.0249)  & ($\pm$0.0267) & ~ \\
        
        TFIDF & 0.6812 & 0.7213 & 0.6981 & 0.6889 & 0.7476 & 0.7462 & 0.7139  & 2  \\
        ~ & ($\pm$0.0227) & ($\pm$0.0226) & ($\pm$0.0137) & ($\pm$0.0211) & ($\pm$0.0373) & ($\pm$0.0253)  & ($\pm$0.0289) & ~ \\ 
        
        Word2Vec & 0.6702 & 0.6937 & 0.6779 & 0.6695 & \textbf{0.7692} & \textbf{0.7687} & 0.7082  & 3  \\
        ~ & ($\pm$0.0242) & ($\pm$0.0154) & ($\pm$0.0175) & ($\pm$0.0162) & ($\pm$0.0264) & ($\pm$0.0233)  & ($\pm$0.0479) & ~ \\ 
        
        LDA & 0.5569 & 0.6459 & 0.6492 & 0.6071 & 0.6928 & 0.6942 & 0.6410  & 4  \\
        ~ & ($\pm$0.0145) & ($\pm$0.0297) & ($\pm$0.0303) & ($\pm$0.0199) & ($\pm$0.0291) & ($\pm$0.0144)  & ($\pm$0.0526) & ~ \\ 
        
        BERTopic & 0.4464 & 0.5951 & 0.5999 & 0.5076 & 0.6045 & 0.6298 & 0.5639  & 5 \\
        ~ & ($\pm$0.021) & ($\pm$0.023) & ($\pm$0.0137) & ($\pm$0.0143) & ($\pm$0.0138) & ($\pm$0.0223) & ($\pm$0.071) & ~ \\
        \hline
    \end{tabular}
    }
\end{table}

Based on the results shown in Table \ref{table:classifiction task}, we first find that the BOW method performs the best, ranking first across all three metrics (Accuracy, AUC, and F1-Score). TF-IDF follows closely in second place, exhibiting stable performance, while Word2Vec ranks third with moderate results. Second, both LDA and BERTopic show significantly lower prediction performance, indicating that topic modeling features may lose discriminative information in supervised classification tasks. Finally, Word2Vec demonstrates greater potential when combined with ensemble learning models (RF and XGBoost), whereas sparse features such as BOW and TF-IDF perform better under traditional machine learning models (LR, ANN, DT, and SVM).

By comparing the regression and classification tasks, we find that representation methods capturing semantic relevance, such as Word2Vec, are better suited for predicting the continuous values of data product prices. In contrast, for price-level classification, representations that ignore semantic relationships, such as BOW and TF-IDF, tend to yield better performance. In both regression-based and classification-based price prediction tasks, topic modeling methods such as LDA and BERTopic exhibit relatively poor performance. In addition, we find that the combination of XGBoost and Word2Vec achieves robust predictive performance in both the classification and regression tasks.

The performance differences between Word2Vec and BOW/TF-IDF can be explained by the fundamental distinctions between regression-based and classification-based pricing tasks. Regression tasks aim to model continuous variations in data product prices and therefore require the model to capture gradual changes in semantic intensity, enabling a smooth mapping between textual semantics and price values. In the context of data product pricing, prices are rarely determined by a single keyword; instead, they emerge from the joint effects of multiple latent semantic dimensions, such as functional richness, data coverage, application scenarios, update frequency, and timeliness. By contrast, classification tasks focus on predicting discrete price levels. Their primary objective is to separate samples into a limited number of categories, emphasizing the identification of clear decision boundaries rather than modeling continuous semantic variation. In such settings, the presence of a small number of highly discriminative keywords is often sufficient to distinguish price levels effectively.

From a representation perspective, Word2Vec maps textual content into a continuous, low-dimensional, and dense semantic space, which makes it particularly well suited for regression tasks. First, data product prices are continuous variables, and the continuity of the Word2Vec semantic space aligns well with this property, facilitating the learning of smooth functional relationships in regression models. Second, semantically similar words are embedded in nearby regions of the vector space, allowing the model to capture the cumulative effect of semantic intensity on pricing while reducing sensitivity to lexical differences. In contrast, BOW and TF-IDF generate high-dimensional and sparse representations that ignore contextual information and semantic similarity, focusing instead on word occurrence and frequency. These characteristics are advantageous in classification tasks, as discriminative keywords linked to specific price tiers are preserved with minimal noise.
\subsection{Feature importance analysis}
\subsubsection{Regression results}
\begin{figure}
    \centering
    \includegraphics[width=1\linewidth]{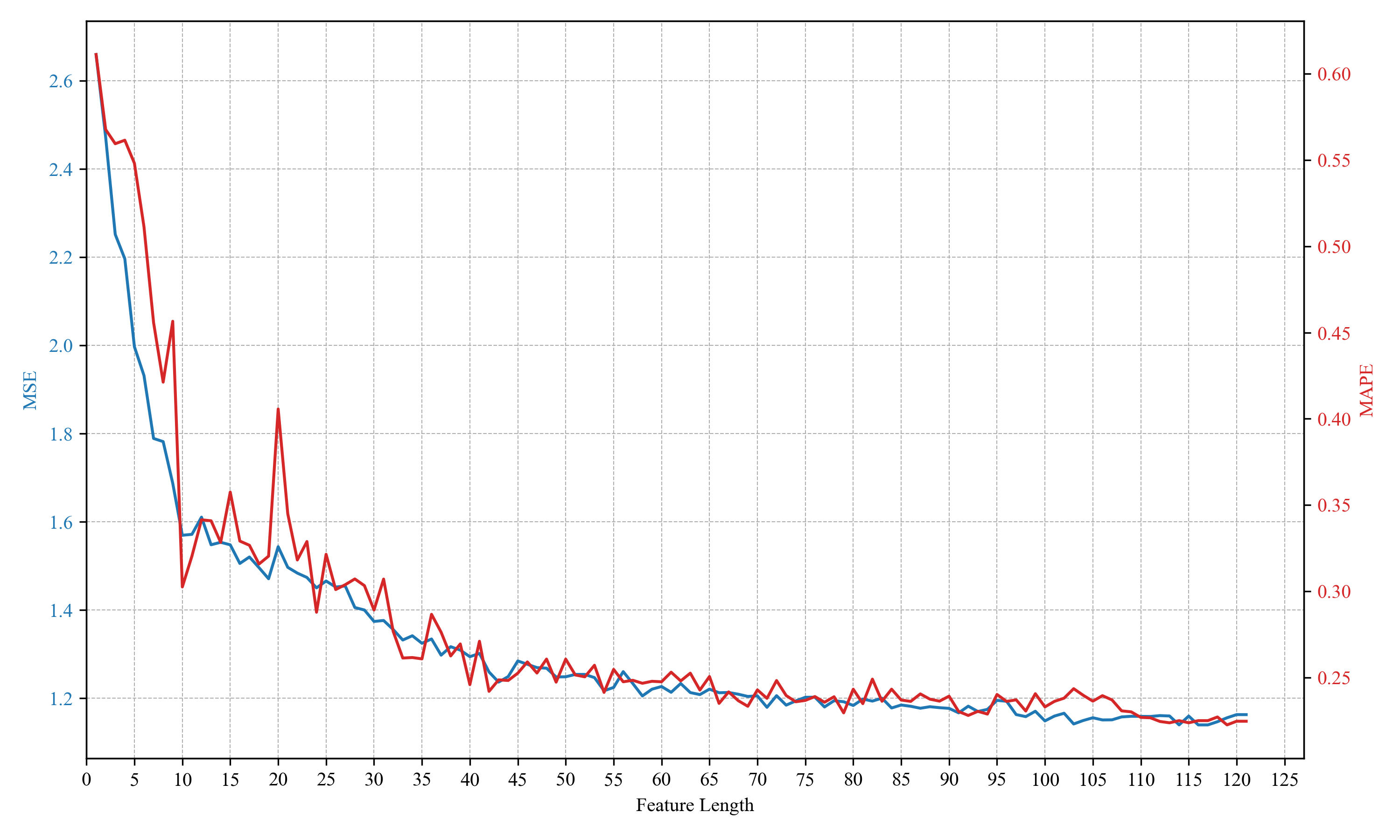}
    \caption{Prediction results based on mRMR method}
    \label{fig:XGBoost-mRMR regression}
\end{figure}

We combine the XGBoost model with the mRMR method to analyze the impact of different feature number on the prediction performance of data product prices. The mRMR method simultaneously considers the relevance between candidate features and the target variable, as well as the redundancy among the selected features. The results of the model are shown in Figure \ref{fig:XGBoost-mRMR regression}.

Based on the results shown in Figure \ref{fig:XGBoost-mRMR regression}, we first find that as the number of features increases from 0 to around 40, both MSE and MAPE decrease rapidly, indicating that adding features at the early stage significantly enhances the model’s prediction performance. When the number of features exceeds approximately 60, both curves stabilize, with MSE around 1.2 and MAPE below 0.25, suggesting that further increasing the number of features provides limited improvement and may even introduce noise. The results confirm the effectiveness of the mRMR method in feature selection, as competitive performance can be obtained with a relatively small number of features (around 40–60).

\begin{figure}
    \centering
    \includegraphics[width=1\linewidth]{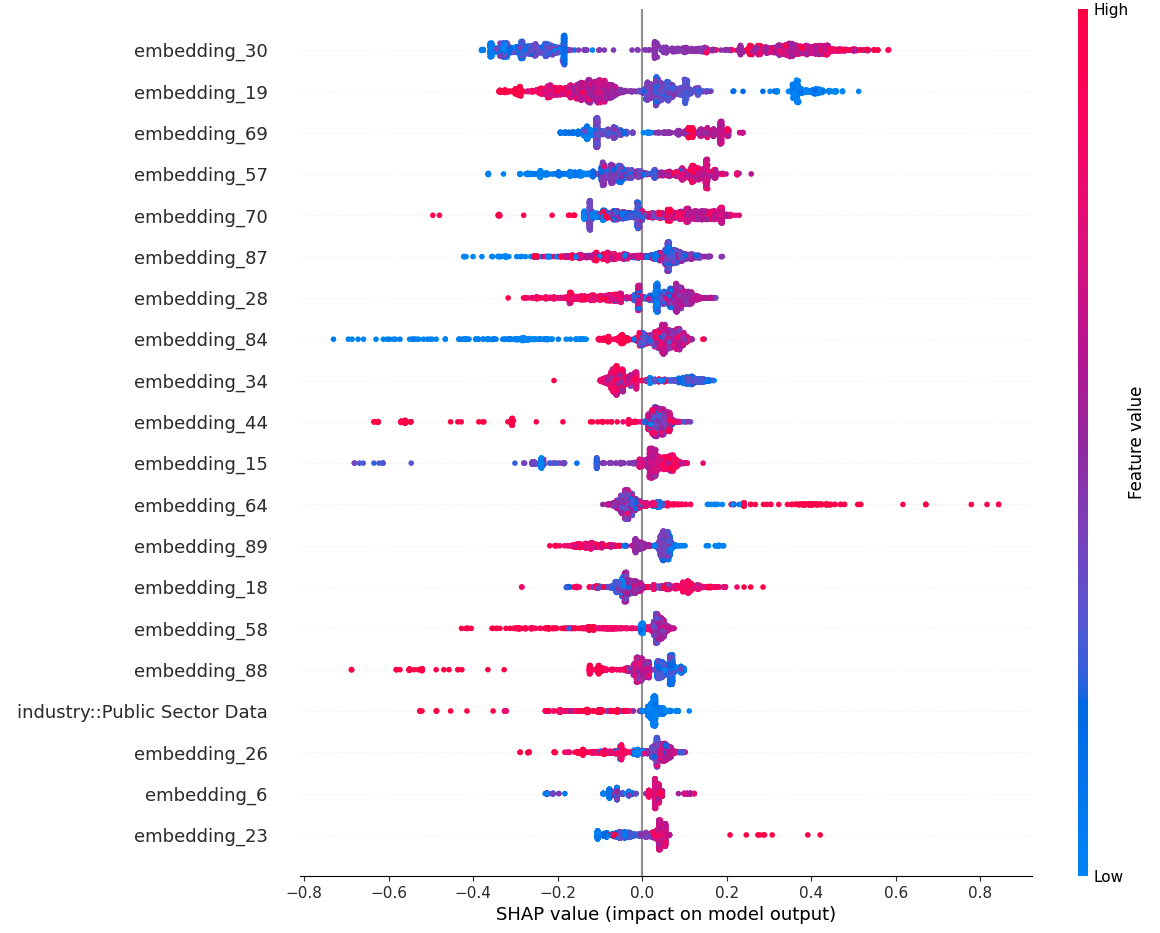}
    \caption{Feature importance ranking (Top 20). In the figure, the color of each data point represents the magnitude of the corresponding feature value: redder points indicate larger feature values, while bluer points indicate smaller values. A positive SHAP value (SHAP value $>$ 0) signifies a positive influence on the output, while a negative SHAP value indicates a negative influence.}
    \label{fig:shap analysis}
\end{figure}

Feature importance analysis can help us understand which factors critically influence data pricing. In this paper, we achieved this goal by utilizing SHAP (SHapley Additive exPlanations), a widely used post-hoc interpretability method. Our calculations yielded the feature importance ranking plot shown in Figure \ref{fig:shap analysis}.

As demonstrated by the results in the figure, the word embeddings generated via text vectorization emerged as the most influential features. Among the top twenty features in the importance ranking, all were word embedding features except for ``industry: Public Sector Data."

Since the numerical values of the word embeddings themselves lack direct semantic meaning, to effectively analyze which specific textual expressions influence data pricing, we identified the 15 words with the highest values and the 15 words with the lowest values along the relevant word vector dimensions for further analysis.

For instance, words with high values along the `embedding\_30' dimension include: ``demographics, gender, medication, therapeutic, age, ...". These expressions are clearly related to population and health, suggesting that data related to healthcare and individual profiles are generally priced higher. Conversely, words with low values in this dimension include: ``ice, weather, days, ...", indicating that data associated with weather or the environment are generally priced lower.

This analysis is consistent with the current reality of data pricing. Due to the sensitivity and difficulty in acquiring healthcare-related data, their pricing is typically higher. In contrast, environmental data are often statistical and less sensitive, leading to comparatively lower prices.

\subsubsection{Classification results}

\begin{figure}
    \centering
    \includegraphics[width=1\linewidth]{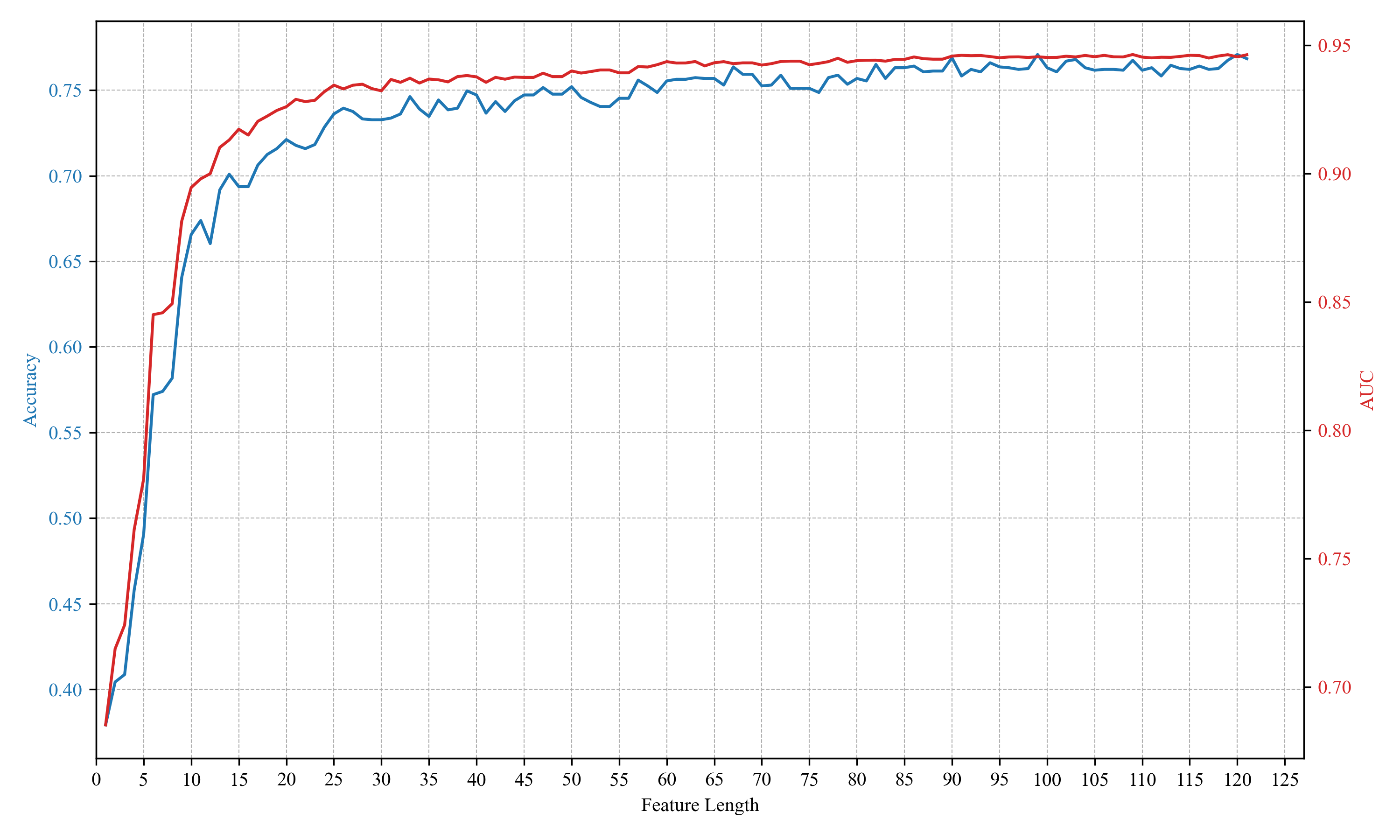}
    \caption{Classification results based on XGBoost+mRMR method}
    \label{fig:XGBoost+mRMR+classification}
\end{figure}

We employ the XGBoost model combined with the mRMR feature selection method to analyze how different numbers of selected features affect the performance of price-range prediction for data products. Accuracy and AUC are used as the evaluation metrics for the classification task. The classification performance of XGBoost+mRMR across varying feature lengths is illustrated in Figure \ref{fig:XGBoost+mRMR+classification}.

Based on the results shown in Figure \ref{fig:XGBoost+mRMR+classification}, we observe that as the number of selected features increases from 0 to approximately 30, both Accuracy (blue curve) and AUC (red curve) exhibit a clear and rapid improvement. In the initial stage (0–15 features), performance rises sharply: Accuracy increases from around 0.40 to approximately 0.70, while AUC improves from about 0.70 to nearly 0.90. The result indicates that the features selected by the mRMR method at this early stage contain high information gain and substantially enhance the model’s discriminatory ability. In the intermediate stage (15–30 features), performance continues to improve but at a slower pace. After approximately 30 features, both metrics enter a stable phase: Accuracy stabilizes around 0.76-0.78, and AUC remains within 0.94-0.95. Adding more features does not provide additional performance gains and may introduce minor noise.

By comparing the classification (Figure \ref{fig:XGBoost+mRMR+classification}) and regression (Figure \ref{fig:XGBoost-mRMR regression}) results, we find that the classification task is easier for the model to separate into distinct categories. Therefore, the highly relevant features selected by mRMR at the early stage quickly improve the model’s ability to distinguish between classes, and the performance becomes stable when around 30 features are included. In contrast, the regression task aims to predict continuous price values, which requires capturing more detailed patterns in the data. As a result, additional features continue to provide incremental benefits, leading to a smoother decline in MSE and MAPE, with performance stabilizing only after about 60 features. This later stabilization point indicates that regression depends on a larger and richer set of features compared with classification.
\begin{figure}
    \centering
    \includegraphics[width=1\linewidth]{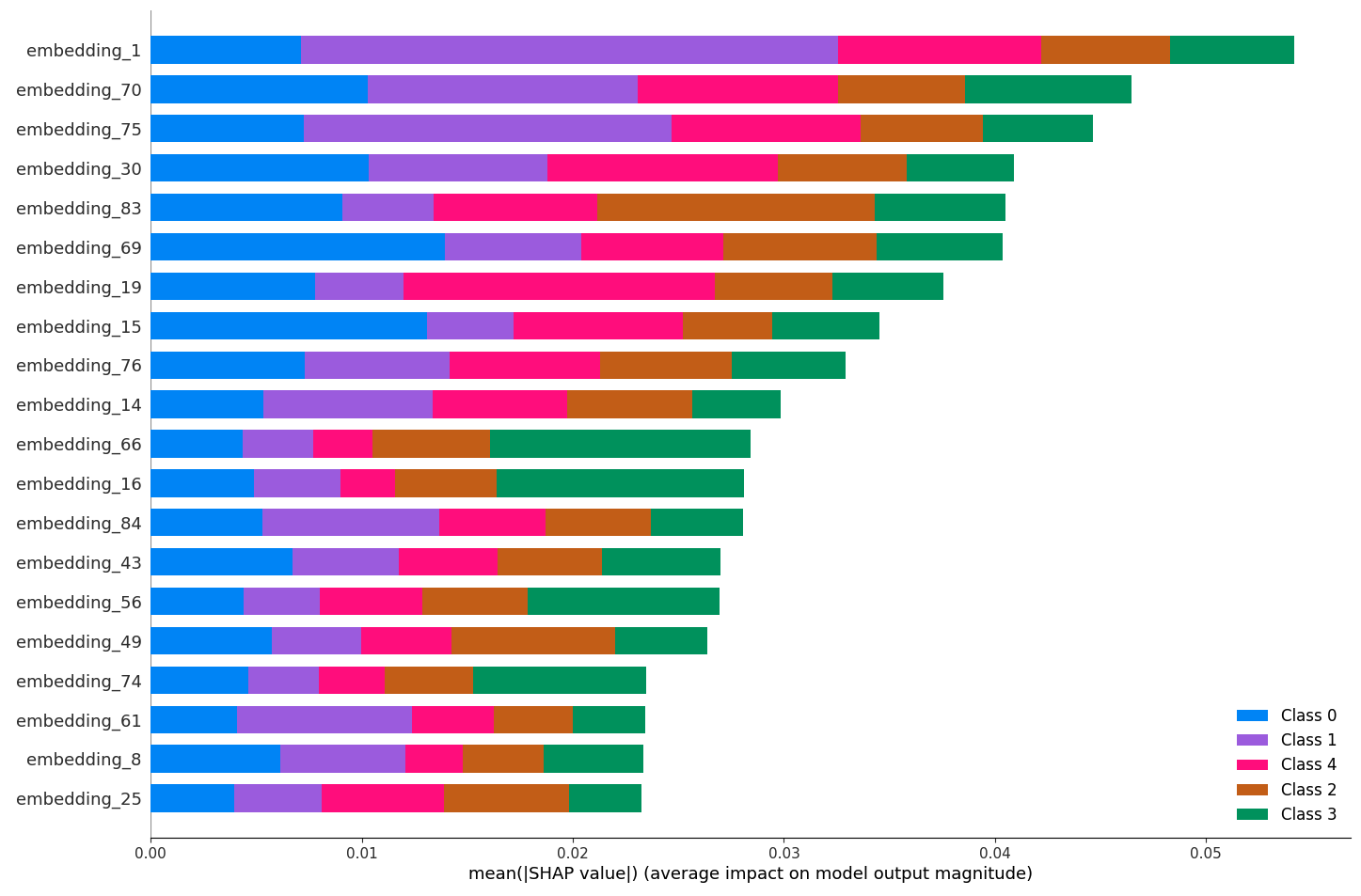}
    \caption{Feature importance ranking of classification model (Top 20)}
    \label{fig:XGBoost-Classification}
\end{figure}

We also present the feature importance ranking for the classification task using SHAP analysis, as shown in Figure \ref{fig:XGBoost-Classification}.
As the results shown in Figure \ref{fig:XGBoost-Classification} indicate, the features important for the classification model and those important for the regression model exhibit both differences and similarities. For example, features such as embedding\_70, embedding\_19, embedding\_30, and embedding\_69 all rank highly in importance across both the classification and regression models, suggesting these features possess general applicability.

The differences are notable: the feature embedding\_1, which is not significant for the regression model, ranks highest in importance for the classification model. Through back-tracing, we determined that embedding\_1 primarily represents macro-level data product descriptions, using terms such as ``national" or ``markets". Consequently, this feature makes a substantial contribution to classifying the lower price category (Class 1).

\section{Discussion} \label{sec:discussion}
\subsection{Theoretical implications}
Compared with the existing studies on data product pricing, this study offers two primary theoretical implications. First, compared with prior works that rely solely on numerical features for pricing \citep{hao2024hceg,lv2025research}, this study extends the exploration of how textual semantic information contributes to data product pricing. Second, in contrast to studies that employ a type of textual representation method \citep{zhu2024dataprice}, we examine how different textual representation techniques influence both continuous price prediction and tiered pricing outcomes. Consequently, this study enriches the literature by broadening the methodological toolkit for textual feature representation and advancing the understanding of pricing mechanisms for data products.

Second, we propose a novel interpretability approach that combines SHAP with a mapping from word embeddings back to specific words. We first use SHAP values to assess the importance of semantic embedding features. Building on the identification of influential word embeddings, we then map these features back to their most representative words by selecting the top-ranked words associated with each embedding dimension. This method allows stakeholders to better understand the underlying drivers of pricing decisions. The method introduced in this study enhances existing interpretability techniques for text-based features and provides a framework for analyzing semantic contributions to data product pricing.

\subsection{Practical implications}
This study offers important managerial implications for the three primary stakeholders in data marketplaces. These implications can be summarized as follows.

First, we find that for the continuous price regression task, Word2Vec, which captures semantic coherence and contextual information, outperforms BOW and TF-IDF. In contrast, for the price-tier classification task, frequency-based methods such as BOW and TF-IDF achieve better performance. These findings suggest that participants in data marketplaces should adopt differentiated textual descriptions strategies when presenting data products. For data providers, using semantically rich and domain-specific textual descriptions can facilitate more accurate price estimation, whereas emphasizing precise and high-frequency keywords is more helpful for positioning products within specific price tiers. For data consumers, when their need is general-purpose, core terms and keyword frequency can serve as indicators of product value. Conversely, for tasks that require customization, carefully reading the full textual description becomes essential for identifying suitable products. For data platform, offering two types of description templates, one tailored for domain-specific data products and another for entry-level products, may improve the efficiency of information display. Additionally, adopting differentiated presentation styles for premium versus basic products can further enhance matching efficiency.

Second, this study identifies which specific features exert significant positive or negative influences on pricing based on SHAP analysis. The results show that semantic features related to medical and demographic information (e.g., embedding\_30) are associated with higher prices, whereas features linked to weather or environmental information tend to correspond to lower price levels. This analytical approach greatly enhances the interpretability of the model’s predictions and supports the development of actionable business insights. For data providers, highlighting the domain-specific high-value embedded in their products may help justify higher prices. For data consumers, such interpretability analyses can clarify the underlying pricing logic of the platform, enabling more cost-effective purchasing decisions. For data platform, presenting feature-level interpretability results prior to purchase can contribute to a more efficient trading environment.

Based on the XGBoost-mRMR model, we find that model performance does not continue to improve as the number of features increases. Instead, optimal or near-optimal prediction performance is achieved with roughly 30 features for the classification task and around 60 features for the regression task. This result suggests that the value of data products is largely concentrated in a limited set of high-information features. For data providers, simply adding more attributes when submitting products does not necessarily raise the price; rather, retaining a concise set of high-information-value features is more beneficial. For data consumers, it is unnecessary to examine every listed attribute in detail; attention should instead focus on the key determinants of value. For data platform, feature-selection techniques such as mRMR can be incorporated into pricing assistance tools to enhance platform efficiency and improve matching quality in data trade.

\section{Conclusion} \label{sec:conclusion}
In this study, we first employ five textual representation methods (Bag-of-Words, TF-IDF, Word2Vec, LDA, and BERTopic) to encode the descriptive text of data products. And then, we employ six different machine learning models, including linear regression, neural networks, decision trees, support vector regression, random forests, and XGBoost, to predict data product prices. Our experimental design includes two tasks: a regression task that predicts the continuous prices of data products, and a classification task that discretizes prices into ordinal levels, enabling a comprehensive comparison of how different textual representation methods perform across distinct prediction scenarios. Finally, we conduct feature importance analysis based on the mRMR feature selection method and SHAP-based interpretability techniques to uncover the key factors that influence data product pricing.

We first find that for the regression task predicting continuous prices, Word2Vec which captures semantic coherence and contextual information outperforms BOW and TF-IDF, whereas in the classification task, BOW and TF-IDF achieve better performance than Word2Vec. In addition, our analysis of the XGBoost model combined with mRMR reveals that model performance does not continually improve as the number of selected features increases. Instead, classification performance reaches its optimal or near-optimal level at around 30 features, while regression performance stabilizes at approximately 60 features. Finally, the SHAP analysis uncovers the specific features that exert significant positive or negative influences on price. We find that semantic features related to healthcare and demographics tend to drive prices upward, whereas those associated with weather and environmental topics are linked to lower prices.

% \singlespacing
% \setlength\bibsep{0pt}
% \bibliographystyle{my-style}

% \bibliographystyle{plainnat}
% \bibliography{reference.bib}

% \clearpage

% \onehalfspacing

% \section*{Tables} \label{sec:tab}
% \addcontentsline{toc}{section}{Tables}

\iffalse
\begin{table}
    \centering
    \caption{Distribution of different data product pricing strategies}
    \label{tab:placeholder}
    \begin{tabular}{cccc}
    \hline
         Data Market & Data pricing strategies & Number & Percentage(\%) \\
    \hline
         Amazon & Time & 2073 & 43.25\\
         & Bytes & 3 & 0.06\\
         & Request & 16 & 0.33 \\ 
         & One-off & 0 & 0 \\
         & Free & 2073 & 43.25 \\
    \hline
        Datarade & Time & 1151 & 25.69\\
        & Bytes & 15 & 0.33\\
        & Request & 148 & 3.30\\
        & One-off & 2945 & 65.74 \\
        & Row & 336 & 7.5\\
        & Usage & 0 & 0 \\
        & Free & 0 & 0 \\
    \hline
    \end{tabular}
    
\end{table}
\fi

% \clearpage

% \section*{Figures} \label{sec:fig}
% \addcontentsline{toc}{section}{Figures}

%\begin{figure}[hp]
%  \centering
%  \includegraphics[width=.6\textwidth]{../fig/placeholder.pdf}
%  \caption{Placeholder}
%  \label{fig:placeholder}
%\end{figure}

\section*{Acknowledgments}
The authors acknowledge the financial support from the National Natural Science Foundation of China (72192802, 72342008, 72401029 and 72473013).

\section*{Declaration of Competing Interests}
The authors declare no competing interests.

\newpage

\appendix

\section*{Appendix}
\subsection*{Appendix A. Prompts}

\begin{promptbox}[Determine refund policy level]
\textbf{Input:} You are a text classification assistant. Your task is to assign a refund policy level (0--4) based on the following rules: 
\begin{itemize}
    \item Level 0: No refunds (clear denial). Examples: "No refunds.", "Refunds are not offered on this product.", "This product is non-refundable.", "Refunds not applicable."
    \item Level 1: Undefined / Not specified. Examples: "This product does not have a defined refund policy.", "Refund policy will be discussed...", "Refunds are not specified for this product."
    \item Level 2: No refunds, but with additional details (trial, sample, disclaimer). Examples: "No refunds. Please utilize trial version before purchase.", "Please request a free sample before buying.", "Not Applicable.", "This is a free sample.", "All sales are final due to digital nature."
    \item Level 3: No refunds, but contact/support is offered. Examples: "No refunds but contact us at ...", "Refunds are not offered, but we will fix issues.", "Please contact support\@... for assistance."
    \item Level 4: Conditional refunds (specific cases allowed). Examples: "Full refund available upon request.", "Refund only if subscription is canceled within 90 days.", "Refunds issued for valid reasons only."
\end{itemize}
Instruction:
Classify each input text into a level 0--4. 
Return the result as a JSON array of integers, in the same order as input texts.
Do not output anything else.

\medskip\hrule\medskip   % 中间的分隔线

\textbf{Output:} [2,0,4,1,3]

\end{promptbox}

\begin{promptbox}[Determine data product industry]
\textbf{Input:} You are a text classification assistant. Your task is to calculate the similarity between the data product description text and each scenario in the given list of application scenario type, with a numerical range of [0,1], where the most similar scenario is assigned a value of 1. Do not output any other content. List of application scenario type and their brief descriptions:
\begin{itemize}
\item E-commerce and Business Data: Data products for e-commerce and online sales, such as sales data, inventory data, consumer behavior data.
\item Retail and Location Data: Data involving geographical location and marketing activities, such as GPS data, advertising data, foot traffic data.
\item Financial Services: Data for the financial industry, such as banking, insurance, investment data.
\item Healthcare and Life Sciences Data: Data related to health, medicine, biology, such as disease data, clinical trial data, genomic data.
\item Resources Data: Data about natural resources, such as energy data, mining data, agricultural data.
\item Public Sector Data: Data from government and public sectors, such as census data, public records, regulatory data.
\item Media and Entertainment Data: Data for media and entertainment, such as streaming data, content rating data, social media data.
\item Telecommunications Data: Data from telecommunications networks and services, such as call records, network performance data.
\item Cars and Automotive Data: Data related to vehicles and transportation, such as sensor data, traffic data, car sales data.
\item Manufacturing Data: Data for manufacturing and industrial processes, such as production data, supply chain data.
\item Environmental Data: Data about the environment, climate, sustainability, such as pollution data, climate indicators.
\item Gaming Data: Data from the gaming industry, such as player statistics, game performance data.
\end{itemize}

Instruction: For N input items, return N lines. For item i, output ONLY a JSON-style list of 12 numbers in [0,1] whose maximum equals 1, corresponding to the scenarios in the exact order listed above. No extra text, no codes fences.

\medskip\hrule\medskip   % 中间的分隔线

\textbf{Example}: Coronavirus (COVID-19) data that has been gathered and unified from trusted sources. This data is provided to the public by Salesforce, MuleSoft, and Tableau at no cost to help you make better decisions, fast.

\textbf{Output:} [0.1, 0.05, 0.05, 1, 0.1, 0.8, 0.05, 0.05, 0.05, 0.05, 0.6, 0.05]

\end{promptbox}

%\subsection*{Appendix B. Tables}
%\setcounter{table}{0}
%\renewcommand{\thetable}{B.\arabic{table}}

\bibliographystyle{apalike}   % 或 aea / econ / chicago
\bibliography{rme-sample}
%\bibliographystyle{utphys}

%\footnote{Ruize Gao, Beijing Institute of Mathematical Sciences and Applications, Beijing, China. 
%Feng Xiao, Beijing Institute of Mathematical Sciences and Applications, Beijing, China. 
%Jinpu Li, Tsinghua University, Beijing, China. 
%Shaoze Cui, Corresponding author, Beijing Institute of Technology, Beijing, China, shaoze-cui@foxmail.com; shaoze-cui@bit.edu.cn}}

\address{Beijing Institute of Mathematical Sciences and Applications (BIMSA)\\
Huairou District, Beijing 101408, China\\
 \email{gaoruize@bimsa.cn}}
 
\address{Beijing Institute of Mathematical Sciences and Applications (BIMSA)\\
Huairou District, Beijing 101408, China\\
 \email{1063710138@qq.com}}
 
 \address{Tsinghua University\\
Haidian District, Beijing 100084, China\\
 \email{li-jp21@mails.tsinghua.edu.cn}} 

 \address{
Beijing Institute of Technology\\
Haidian District, Beijing 100081, China\\
 \email{shaoze-cui@foxmail.com; shaoze-cui@bit.edu.cn}\\
 \received{November 27, 2025}\\
 \accepted{February 12, 2026}}

\end{document}